\documentclass[runningheads]{llncs}
\usepackage[T1]{fontenc}
\usepackage[colorlinks=true,urlcolor=teal,
            linkcolor=teal,citecolor=teal]{hyperref}
\usepackage{tabularx}
\usepackage{booktabs}
\usepackage{array}
\usepackage{xspace}
\usepackage{amsmath}
\usepackage{amssymb}
\usepackage{xcolor}
\usepackage{caption}
\usepackage{comment}
\usepackage[expansion,protrusion]{microtype}
\usepackage[backend=biber,
  maxcitenames=3, minbibnames=1, mincitenames=1, maxbibnames=3,
  uniquelist=false, giveninits=true, url=false, doi=false,
  eprint=false, isbn=false, style=lncs]{biblatex}
\usepackage{listings}
\usepackage[frozencache, cachedir=.]{minted}
\setminted{fontsize=\scriptsize}
\definecolor{LightGray}{gray}{0.95}
\usepackage{graphicx}
\usepackage{subcaption}
\usepackage[shortcuts]{extdash}
\usepackage{tikz}
\usepackage[export]{adjustbox} 
\usetikzlibrary{arrows.meta,arrows,positioning,calc,shadows.blur,fit,patterns,shapes.arrows,shapes.geometric}
\usepackage{color}

\newcommand{\reprourl}{https://github.com/lfd/Quantum_Experiment_Framework}
\newcommand{\zenodourl}{https://zenodo.org/records/17524220}

\usepackage{censor}
\newboolean{anonymous}
\setboolean{anonymous}{false}
\ifbool{anonymous}{}{}
\ifbool{anonymous}{}{\renewcommand{\blackout}[1]{#1}}
\ifbool{anonymous}{\renewcommand{\reprourl}{https://github.com/anonym789/Quantum_Experiment_Framework}}{}
\ifbool{anonymous}{\newcommand{\genemail}[1]{\email{xxx.xxx@xx.xx}}}{\newcommand{\genemail}[1]{\email{#1}}}

\newcommand{\ie}{\emph{i.e.}\xspace}
\newcommand{\eg}{\emph{e.g.}\xspace}

\newcommand{\etal}{\emph{et al.}\xspace}

\newbibmacro{arxiv}[1]{%
  \iffieldundef{archivePrefix}{arXiv}{\thefield{archivePrefix}}:%
  \iffieldundef{eprint}{}{\thefield{eprint}}
}
\DeclareFieldFormat*{title}{\usebibmacro{string+doi}{\mkbibemph{#1}}}

\DeclareBibliographyDriver{misc}{%
  \usebibmacro{bibindex}%
  \usebibmacro{begentry}%
  \printnames{author}%
  \setunit*{\labelnamepunct}\newblock%
  \usebibmacro{title}, %
  \newunit\newblock
  \usebibmacro{arxiv}%
  \newunit\newblock
  \printfield{addendum}%
  \setunit{\bibpagerefpunct}\newblock
  \usebibmacro{pageref}%
  \newunit\newblock
  \iftoggle{bbx:related}
  {\usebibmacro{related:init}%
    \usebibmacro{related}}
  {}%
  \usebibmacro{doi+eprint+url}%
  \nopunct%
  \printfield{year}
  \nopunct%
  \usebibmacro{finentry}%
}

\begin{document}
\title{QEF: Reproducible and Exploratory\\ Quantum Software Experiments}

\author{\blackout{Vincent Gierisch\inst{1}\orcidID{0009-0004-0638-6757} \and\\
Wolfgang Mauerer\inst{1,2}\orcidID{0000-0002-9765-8313}}}
\authorrunning{\blackout{Gierisch and Mauerer}}
\institute{\blackout{Technical University of Applied Science Regensburg, Germany}\\
\genemail{vincent.gierisch@othr.de},
\genemail{wolfgang.mauerer@othr.de} \and
\blackout{Siemens AG, Foundational Technologies, 
Munich, Germany}}
\maketitle
\begin{abstract}
Commercially available Noisy Intermediate-Scale Quantum (NISQ) devices now make small hybrid quantum–classical experiments practical, but many tools hide configuration or demand ad-hoc scripting.

We introduce the Quantum Experiment Framework (QEF): A lightweight framework designed to support the systematic, hypothesis-driven study of quantum algorithms. 
Unlike many existing approaches, QEF emphasises iterative, exploratory analysis of evolving experimental strategies rather than exhaustive empirical evaluation of fixed algorithms using predefined quality metrics. 
The framework’s design is informed by a comprehensive review of the literature, identifying principal parameters and measurement practices currently reported in the field.

QEF captures all key aspects of quantum software and algorithm experiments through a concise specification that expands into a Cartesian product of variants for controlled large-scale parameter sweeps. 
This design enables rigorous and systematic evaluation, as well as precise reproducibility. 
Large sweeps are automatically partitioned into asynchronous jobs across simulators or cloud hardware, and ascertain full hyper-parameter traceability. 
QEF supports parameter reuse to improve overall experiment runtimes, and collects all metrics and metadata into a form that can be conveniently explored with standard statistical and visualisation software.

By combining reproducibility and scalability while avoiding the complexities of full workflow engines, QEF seeks to lower the practical barriers to empirical research on quantum algorithms, whether these are designed for current NISQ devices or future error-corrected quantum systems.

\end{abstract}

\section{Introduction}
Quantum computing seeks to harness the computational power of microscopic systems to accelerate calculations for generic or specific problems over classical competitors.
Current physical implementations, commonly referred to as noisy intermediate-scale quantum (NISQ) systems, remain limited in size and operational quality. They are too small to execute any of
the known algorithms with mathematically provable exponential (or other) advantage.
Yet, they support hybrid quantum-classical approaches such as variational quantum circuits (VQCs). Properties and behaviour of such heuristics have received intense discussion in literature, yet their potentials, performance and limitations are not yet fully understood, and they are often evaluated and benchmarked
empirically~\cite{yue:2023:qsa}. 
While this is not along the traditional lines of reasoning in computer science, it has become an accepted standard practice not only in quantum computing, but also in other fields like machine learning, where an incomplete a-priori theoretical understanding of algorithmic capabilities often prevails. 
Given the many challenges of empirical deduction, it is a crucial problem of how to perform the required characterisation of quantum methods. 
Likewise, it is important to emphasise that the problem not only applies to the NISQ era, but will pervade into the era of fully error-corrected quantum computers and sound algorithms that enjoy a solid complexity-theoretic understanding, as a 
characterisation of their systemic and workload-specific
properties will likely remain a challenge that can only be met empirically, given the complex interplay of many quantum and classical components.

When developing and testing such quantum algorithms to solve specific problems, researchers can rely on a wide range of supporting frameworks.
These tools often make it particularly easy to rapidly design quantum algorithms.
Some frameworks even offer interfaces that allow immediate execution of the developed algorithm on quantum simulators or real quantum hardware. 
However, they differ significantly in the level of control and flexibility they provide researchers.
While some provide only basic components for building parametrised quantum circuits, others can generate entire algorithms based almost solely on the description of the combinatorial problem.
However, this diversity comes at a cost: many existing tools either demand extensive ad-hoc scripting or restrict researchers’ flexibility. 
As a result, they are often unsuited for hypothesis-driven experimentation, where the effect of individual parameters must be isolated in a controlled and reproducible way.

\begin{figure}[htb]
    \centerline{\input{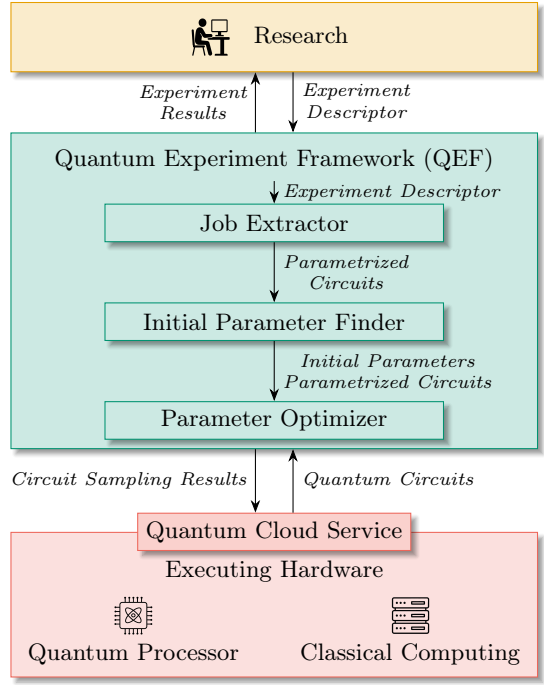}}
    \caption{\label{fig:QLMIntro}
    Integration of QEF into research workflows. 
    Experiments defined via a structured configuration expand into a Cartesian product of job variants.
    These jobs are sent to the target hardware, where both, quantum processors and classical computing resources, collaborate to run and evaluate hybrid algorithms. 
    Result data are stored in formats optimised for convenient and dynamic exploratory and visual analysis with standard tools.}
\end{figure}

To address this gap, \emph{Quantum Experiment Framework} (QEF) 
provides a framework for systematic and reproducible studies of hybrid quantum-classical algorithms (see an overview in \autoref{fig:QLMIntro}).
It streamlines quantum algorithm and software research through four key aspects:

(1) QEF uses a simple, structured configuration to describe experiments that fully captures every supported parameter. 
This eliminates ad-hoc scripting and enables consistent exploratory analysis.
(2) It automatically splits large studies into asynchronous jobs and dispatches them aptly in parallel.
This enables large parameter sweeps without adding manual overhead for researchers.
(3) QEF supports built-in parameter reuse between runs, which can accelerate convergence when exploring sequences of related experiments~\cite{zhou:2020:physicalreviewx}. 
(4) QEF logs all results and metadata in tidy format~\cite{wickham:2014:jss}.
This gives researchers immediate access to metrics collected at every iteration step of a hybrid quantum-classical experiment.

We believe this makes QEF a practical and researcher-centric framework to ease hypothesis-driven studies of quantum algorithms and software.
We provide a reproduction package~\cite{Mauerer:2022} for our framework,
including the case study, in the form of a \href{\reprourl}{code repository} and \href{\zenodourl}{reproduction package}.

\section{Related Work}
Most available quantum computing frameworks, such as Qiskit or Pennylane, do not focus on experiment preparation and result data collection. 
Nevertheless, frameworks exist that are designed with a particular focus on specific components of a hybrid quantum-classical experiment workflow. In this section, we review
related work, but also elaborate on how the design of existing tools differs from the design choices made for QEF.

Many approaches focus on the creation and execution of hybrid quantum-classical algorithms~\cite{bergholm:2022:pennylane, broughton:2021:tensorflow, sharma:2022:openqaoa, johnson:2022:qiskitruntime, hooyberghs:2021:azure, mauerer:2005:cqpl,carbonelli:24}. However, most do not focus on dynamic workflow orchestration and experimentation like QEF. 
Based on their documentation and example repositories, reproducing QEF's experiment workflow requires user-written code for generating parameter grids,
launching batches across back-ends, monitoring jobs, as well as collecting and storing results.

\subsection{Workflow Orchestration and Experiment Frameworks}
Running and scaling hybrid quantum-classical experiments requires tools that can organise, execute, and track experiments across both quantum and classical systems. 
There are already some frameworks with that aim, offering features for managing workflows and handling experiment data more effectively.

Sivarajah~\etal introduced \href{https://github.com/CQCL/tierkreis}{Tierkreis} in 2022, a framework for hybrid quantum-classical computing~\cite{sivarajah:2022:tierkreis}.
It represents programs as dataflow graphs, allowing developers to build complex hybrid workflows. 
Sauerwein~\etal integrated \href{https://aws.amazon.com/blogs/quantum-computing/tracking-quantum-experiments-with-amazon-braket-hybrid-jobs-and-amazon-sagemaker-experiments/}{Amazon Braket} with \href{https://aws.amazon.com/de/blogs/quantum-computing/tracking-quantum-experiments-with-amazon-braket-hybrid-jobs-and-amazon-sagemaker-experiments/}{SageMaker} in 2022 to systematically track metrics, hyper-parameters, and experiment result data across
multiple different runs. 
Unlike QEF it is tightly coupled to Amazon Braket and does need ad-hoc scripting to run experiments.

Quokka~\cite{beisel:2022:quokka}, introduced by Beisel~\etal in 2023, decomposes hybrid algorithms into microservices orchestrated by an external workflow engine.
In contrast to Quokka, QEF centres the experiment; it not only orchestrates quantum algorithms but also enables running systematic studies on them.
Quokka, by comparison, targets the application as a workflow of services. 
It decomposes a typical variational quantum algorithm (VQA) loop into microservices, like circuit generation, execution, readout, error mitigation, objective evaluation, and optimisation.
These services are intended to be driven by an external workflow engine.
This service-oriented style emphasises loose coupling and interchangeability of single microservices.
In short, QEF focuses on the experiment with a hybrid quantum-classical algorithm, 
while Quokka concentrates on the components of the hybrid quantum-classical algorithm itself and their integration into a workflow. 
Consequently, QEF aims to support end-to-end experiment creation, batching and result consolidation, whereas Quokka optimises the decomposition and substitution of individual workflow stages.
The two frameworks could complement each other, with QEF extending Quokka to handle the creation and management of experiments.

\subsection{Benchmarking and Evaluation Frameworks}
Benchmarking a multitude of aspects in quantum computing has received considerable methodological attention~\cite{lorenz:2025:benchmarkingquantum}.
QEF is suitable for application-level benchmarking, that is, comparing the performance of quantum algorithms.
By design, it is not intended to target any component-level benchmarks that concern the hardware implementation, nor any system-level benchmarks
It likewise does not support hpc-level benchmarking, as key metrics (\eg, queue times, cloud overhead) are not recorded.
Software-level benchmarking is also outside QEF's scope, since the necessary metrics are not collected.

\href{https://github.com/QUARK-framework/QUARK}{QUARK}~\cite{finvzgar:2022:quark} was proposed in 2022 by Finžgar~\etal as an application-level benchmarking tool for quantum computing. 
It provides an environment to design, implement, execute, and analyse quantum application benchmarks in a reproducible manner. 
Instead of focusing only on low-level metrics like gate fidelity or circuit depth, QUARK intends to facilitate end-to-end comparisons on industrially relevant tasks, such as optimisation problems in logistics, by running complete hybrid workflows and collecting solution quality, runtime, and resource metrics. 
QUARK allows testing one algorithm variant at a time and outputs a result report containing predefined quality metrics.
It is therefore well-suited to test hypotheses about which approach or back-end performs best on realistic tasks.
QEF, in contrast, allows testing many variants simultaneously, thereby isolating the impact of one factor, and returning the results in a tidy data format.

Another benchmarking effort, \href{https://superstaq.readthedocs.io/en/stable/apps/supermarq/supermarq.html}{SupermarQ}~\cite{tomesh:2022:supermarq} introduced by Tomesh~\etal, aims to evaluate quantum processors, 
but it emphasises static circuit families rather than hybrid loops.

In contrast to general-purpose hybrid frameworks that require complex configurations, QEF aims to support exploratory analysis with frequently changing parameters.
It fills the gap between application-level benchmarking frameworks and vendor-specific services as a lightweight, research-oriented experimentation framework for hypothesis-driven studies that yields uniform results.

\section{Literature-Driven Requirements for Quantum Experiments}
\label{sec:empirical_data_collection}
To align the capabilities of QEF with the requirements of the quantum software research community, we performed a systematic
literature review following Kitchenham~\etal~\cite{kitchenham:2009:infsoftwtechnol}.
To balance comprehensiveness, generalisability, and required effort, we consider around 30 studies published in IEEE QCE 2024.\footnote{The following publications have been used:
\href{https://doi.org/10.1109/QCE60285.2024.10278}{10278}, 
\href{https://doi.org/10.1109/QCE60285.2024.10290}{10290}, 
\href{https://doi.org/10.1109/QCE60285.2024.10300}{10300}, 
\href{https://doi.org/10.1109/QCE60285.2024.10246}{10246}, 
\href{https://doi.org/10.1109/QCE60285.2024.10284}{10284}, 
\href{https://doi.org/10.1109/QCE60285.2024.10288}{10288}, 
\href{https://doi.org/10.1109/QCE60285.2024.10268}{10268}, 
\href{https://doi.org/10.1109/QCE60285.2024.10299}{10299}, 
\href{https://doi.org/10.1109/QCE60285.2024.10291}{10291}, 
\href{https://doi.org/10.1109/QCE60285.2024.10318}{10318}, 
\href{https://doi.org/10.1109/QCE60285.2024.10263}{10263}, 
\href{https://doi.org/10.1109/QCE60285.2024.10295}{10295}, 
\href{https://doi.org/10.1109/QCE60285.2024.10271}{10271}, 
\href{https://doi.org/10.1109/QCE60285.2024.10289}{10289}, 
\href{https://doi.org/10.1109/QCE60285.2024.10248}{10248}, 
\href{https://doi.org/10.1109/QCE60285.2024.10264}{10264}, 
\href{https://doi.org/10.1109/QCE60285.2024.10266}{10266}, 
\href{https://doi.org/10.1109/QCE60285.2024.10265}{10265}, 
\href{https://doi.org/10.1109/QCE60285.2024.10245}{10245}, 
\href{https://doi.org/10.1109/QCE60285.2024.10325}{10325}, 
\href{https://doi.org/10.1109/QCE60285.2024.10337}{10337}, 
\href{https://doi.org/10.1109/QCE60285.2024.10338}{10338}, 
\href{https://doi.org/10.1109/QCE60285.2024.10356}{10356}, 
\href{https://doi.org/10.1109/QCE60285.2024.10357}{10357}, 
\href{https://doi.org/10.1109/QCE60285.2024.10361}{10361}, 
\href{https://doi.org/10.1109/QCE60285.2024.10369}{10369}, 
\href{https://doi.org/10.1109/QCE60285.2024.10376}{10376}, 
\href{https://doi.org/10.1109/QCE60285.2024.10388}{10388}. 
Each number completes the DOI https://doi.org/10.1109/QCE60285.2024.\(\langle\)XXXXX\(\rangle\).} 
As QCE counts among the major events for quantum software research, we are confident that our choice of scope does not unduly distort results.

We selected empirical studies of quantum algorithms, in particular, hybrid quantum-classical algorithms. Inclusion criteria require that a study details both input parameters of the experiment (\eg, problem size, circuit configuration, optimiser choices) as well as observed results. We exclude purely theoretical papers without empirical evaluation and studies that lack quantitative experimental results.
This process yielded nearly thirty relevant studies with topics such as variational algorithm benchmarking, quantum circuit optimisation, and noise-aware algorithm design.
From these, a set of common parameters and metrics was extracted. We retained observables and parameters reported by at least two independent studies. \autoref{tab:params_metrics_support} consolidates these frequently encountered parameters and metrics, indicating the extent of their support in QEF.

\begin{table*}[htbp]
    \centering
    \caption{
        Parameters and metrics obtained from the literature review and usage frequency across studies (see the text for QEF column symbol definitions).
    }\label{tab:params_metrics_support}
    
    \begin{tabular}{lcr@{\hspace*{0.5em}}|@{\hspace*{0.5em}}lcr}
        \toprule
        \multicolumn{3}{c@{\hspace*{0.5em}}|@{\hspace*{0.5em}}}{\textbf{Algorithmic/Problem Parameters}} &
        \multicolumn{3}{c}{\textbf{Evaluation Metrics}} \\
        \emph{Parameter} & \emph{QEF} & \emph{Freq.} &
        \emph{Metric}    & \emph{QEF} & \emph{Freq.}\\
        \midrule
            QAOA Layers               & \checkmark   & 37\% & Circuit Depth            & \checkmark    & 37\% \\
            VQA Ansatz Depth          & \checkmark   & 37\% & Approximation Ratio      & (\checkmark)  & 26\% \\
            Custom VQA Ansatz         & \checkmark   & 33\% & Convergence Behaviour    & \checkmark    & 26\% \\
            Custom QAOA Circuit       & \checkmark   & 33\% & Algorithm Runtime        & \checkmark    & 22\% \\
            Problem Size              & \checkmark   & 22\% & Variance in Results      & (\checkmark)  & 19\% \\
            Classical Optimisation Method & \checkmark  & 19\%  & Circuit Height       & \checkmark    & 11\% \\
            Target Back-end            & \checkmark   & 15\% & Total Number of Gates   & \checkmark    & 11\% \\
            Noise Model               & \checkmark   & 15\% & Scalability Behaviour    & (\checkmark)  & 11\% \\
            Initialisation Parameters & \checkmark   & 15\% & Number of Specific Gates & \checkmark    &  7\% \\
            Compilation Schema        & (\checkmark) &  7\% & State Fidelity           & (\checkmark)  &  7\% \\
            Mitigation Method         & $\times$     &  7\% & Feasibility of Solution  & (\checkmark)  &  7\% \\
            Learning Rate             & $\times$     &  7\% & Runtime of Sub-steps     & $\times$      &  7\% \\
            Measurement Shots         & \checkmark   &  7\% & Schedule Data Size       & $\times$      &  7\% \\
            Training Budget           & $\times$     &  7\% & Number of Iterations     & \checkmark    &  7\% \\
                                      &              &      & Cost Function Evaluations     & \checkmark    &  7\% \\
                                      &              &      & Per-iteration Runtime    & \checkmark    &  7\% \\
                                      &              &      & Noise Robustness         & (\checkmark)  &  7\% \\
    \bottomrule
    \end{tabular}
\end{table*}

In \autoref{tab:params_metrics_support}, a check mark $\checkmark$ denotes full native support in QEF, 
while a parenthesised check ($\checkmark$) indicates that the item is not directly supported but can be derived from other data QEF collects.
A cross $\times$ marks parameters or metrics unsupported by QEF (\ie, not recorded and not inferable from other logged data).
To keep the framework lightweight, QEF natively supports only the most commonly used parameters and metrics.
Less common ones are deliberately not included; if needed, they must be implemented on a per-experiment basis (\eg, via experiment-specific extensions or custom collectors).
By limiting native support to widely used parameters and metrics, QEF reduces core complexity and overhead, while still enabling rare needs via experiment-specific extensions. 

The left-hand side of \autoref{tab:params_metrics_support} lists \emph{algorithmic and problem parameters} that characterise the experimental configuration.
These include fundamental aspects such as problem characteristics, configuration of the classical optimiser in a hybrid quantum-classical algorithm, and circuit depth. 
It also includes structural features like the number of qubits (\ie, circuit height), gates, and circuit layers. 
In addition to setting options for the execution platform, such as the back-end and noise model.
For the compilation schema, QEF provides support for all predefined optimisation stages provided by Qiskit, as these are the stages used in all the studies.
The learning rate parameter and the training budget (\ie, maximal training time and maximal epochs) are not natively supported and have to be treated by the researcher per experiment.
The same applies to error mitigation.
Many established methods are problem- and hardware-specific, requiring substantial configuration and tuning.
Other techniques (\eg, zero-noise extrapolation, probabilistic error cancellation) introduce significant complexity and require closer coupling with specific hardware or simulator capabilities.
To keep QEF's core lightweight, we deliberately do not provide native implementations of error mitigation.
Experiments involving mitigation can still be performed with QEF, implemented either as preparatory calibration/compilation steps or as post-processing of the recorded data.

The right-hand side of \autoref{tab:params_metrics_support}  outlines \emph{evaluation metrics} used to assess algorithm performance. 
These range from structural indicators, such as circuit depth and height, the total number of gates in the circuit, the number of specific gates in the circuit (\eg, number of two-qubit gates),
to performance metrics, including total runtime, fidelity, and approximation ratio, as well as behavioural analysis tools, including convergence behaviour and schedule data size.
Here, QEF fully supports a subset of metrics, including algorithm runtime, circuit dimensions, loss function tracking, and gate counts.
Other measurements, such as state fidelity, solution feasibility, and approximation ratio, can be derived from supported metrics.

Informed by the review, we implemented QEF as a lightweight, broadly applicable framework that provides native support for commonly used parameters and metrics,
while intentionally leaving rare, specialised needs to per-experiment extensions rather than expanding the core.

\section{Quantum Experiment Framework} 
In this section, we discuss architecture, functionality, and key features of our framework. 
We tested and studied a prototypical implementation on the Qaptiva 800 platform, which is a high-performance computing appliance that can simulate
noisy and noiseless quantum systems. While our prototype currently only 
supports this single back-end, we plan on adding support for other
targets and simulators.

\subsection{Descriptor \& Results} 
A structured configuration file describes the experiment, and results are systematically recorded. We describe both artefacts in the following.

\subsubsection{Experiment Configuration File}
A single JSON configuration file defines each experiment.
It provides an unambiguous, structured specification of all parameters, 
enabling the framework to parse the experiment's configuration and orchestrate runs (see \autoref{fig:caseStudyExcerpt}).
It comprises four sections:
(1) job configuration, which defines the problem and its solution approach;
(2) optimisation configuration, which sets the optimisation method;
(3) initialisation configuration, which determines how initial parameters are chosen; and
(4) back-end configuration, which specifies the execution back-end for the quantum algorithm.

The job configuration covers the subject problem, which can be defined by type, size, and 
circuit ansatz that should be used. 
It is also possible to define a custom circuit with QASM3~\cite{cross:2022:acm} together with an observable.
QEF also supports different predefined problems, for instance, Max-Cut and Travelling Salesman, as well as custom 
Quadratic Unconstrained Binary Optimisation (QUBO) formulations, as they can have a significant impact on experimental outcome~\cite{gogeissl:2024:qdata}.
As ansatz, QAOA and VQA can be chosen.

The optimisation configuration includes the numerical optimiser and its specific options, as well as the number of shots for the cost function and the gradient evaluation.
The initialisation configuration is determined by strategy, the identifier of a preceding job whose optimised parameters serve as initial guess, and the number of randomly initialised jobs to execute a grid search.

The final section in the configuration file involves configuring the executing back-end. 
Since hardware properties such as qubit connectivity and native gate sets strongly impact circuit depth and overall scalability, QEF’s back-end abstraction allows the injection of hardware-aware configurations \cite{safi:2023:qsw}.
Firstly, the QPU to use can be specified. 
In our implementation, we focused on two QPUs: a noiseless QPU and a noisy QPU.
For the latter, it is possible to provide a noise model that precisely describes the errors encountered in gate operations and measurements. 
To simulate the constraints of a back-end, a topology and connectivity graph can be specified, along with the back-end gate set. 

This configuration file defines the entire experiment and its execution environment, making it easy to reproduce experiments.

\subsubsection{Experiment Results}
QEF retrieves completed jobs from the back-end platform asynchronously.
Results are stored in CSV files in tidy data format~\cite{wickham:2014:jss}, which uses analysis-ready tables where each variable is a column,
each observation (\eg, iteration) is a row, and each table captures one observational unit.
This minimises ad-hoc reshaping and supports standard tooling for visualisation and modelling.
The outcome of each experiment is stored in separate directories.
Notably, two files contain the most significant information.
The first file contains hyper-parameters for the particular experiment, while the second captures the experiment's result. 

Apart from setting experimental parameters, the hyper-parameter file also 
includes an optimal solution to the problem instance (obtained by iterating through all possible states and calculating the costs) to gauge the proximity to optimality,
together with the probabilities and cost values associated with the quantum algorithm’s outcomes.
The result file captures not only the results of the quantum algorithm, like the expectation value, the state vector, and the parameters, but also the outcome of each optimisation iteration. 

\subsection{Architecture \& Workflow}
From an experiment descriptor, QEF derives all parameter combinations as variants, each executed as a separate job composed of tasks representing the individual execution steps.
Independent jobs allow each hypothesis to be tested in parallel, and isolate its impact.
Initial circuit parameters are chosen either by using prior runs as warm start or by a given initialisation strategy.

For evaluation, QEF retrieves or computes the brute-force optimum to enable approximation ratio analysis, recording it alongside all job identifiers and hyper-parameters. It then assembles the execution stack~--~comprising  \emph{InitHandler}, \emph{AllroundOptimizer}, and \emph{QPU}~--~and submits the prepared jobs.
Results can be retrieved later, reflecting common cloud usage patterns, as quantum jobs may require extended runtimes on queued or cloud-based back-ends.

\begin{figure}[htb]
    \centering
    \begin{minipage}{.58\textwidth}
        \centering
        \definecolor{orange_lfd}{HTML}{E69F00}
\definecolor{green_lfd}{HTML}{009371}
\definecolor{purple_lfd}{HTML}{beaed4}
\definecolor{red_lfd}{HTML}{ed665a}

\begin{tikzpicture}[every node/.style={semithick,align=center,
                    font=\scriptsize, 
                    text centered,text width=2cm,minimum height={0.5cm}},
                    node distance=4mm and 0.5cm,
                    shadow/.style={blur shadow}
]

    \node[draw,dashed] (client) {Client Machine};
    \node[draw,dashed, right = of client] (server) {Target};

    \node[draw, below = of client] (job) {Job generation};
    \node[draw, below = of job] (data) {Metadata base};
    \node[draw, below = of data] (stack) {Execution stack build};
    \node[draw, right = of stack] (calc) {(Parallel,hybrid) processing};
    \node[draw, dashed, below = of stack] (idle) {Coffee Break};
    \node[draw, below = of idle] (extract) {Job extraction};
    \node[draw, below = of extract] (post) {Postprocessing};
    \node[draw, below = of post] (base) {Database};
    \node[right = of base] (target_end) {};

    \draw[-Stealth] (stack) -- (calc);
    \draw[-Stealth] (job) -- (data);
    \draw[-Stealth] (data) -- (stack);
    \draw[-Stealth] (calc) |- (extract);
    \draw[-Stealth] (extract) -- (post);
    \draw[-Stealth] (post) -- (base);
    \draw(data) -- ++(-1.5,0) |- (extract)[-Stealth];

    \node[draw=orange_lfd,dotted,fit=(client) (job) (data) (stack) (idle) (extract) (post) (base), very thick] (quark3){};
    \node[draw=orange_lfd,dotted,fit=(server) (calc) (target_end), very thick] (quark4){};

    \node[draw,dashed, left = 7mm of job, text width = 7mm] (datain) {\rotatebox{90}{\begin{tabular}{c}Experiment\\Configuration\end{tabular}}};
    \node[draw,dashed, left = 7mm of base, text width = 7mm] (dataout) {\rotatebox{90}{Evaluation}};

    \draw[-Stealth] (datain) -- (job);
    \draw[-Stealth] (base) -- (dataout);
\end{tikzpicture}
        \captionsetup{width=0.95\linewidth}
        \caption{\label{fig:QLMOverview}QEF workflow. 
        Experiments are expanded into jobs that run asynchronously.
        Completed jobs can be retrieved for result analysis.
        }
    \end{minipage}\hfill
    \begin{minipage}{.38\textwidth}
        \centering
        \vspace{1.1cm}
        \begin{minted}[bgcolor=LightGray]{json}
{
    "job_config_searchspace": {
        "problem_type": ["MaxCut"],
        "nqubits": [10],
        "var_ansatz": ["QAOA"],
        "var_layers": [3, 4, 5],
        "problem_seed": [1],
        "p_edge": [0.7]
    },
    "opt_config_searchspace": {
        "method": ["COBYLA", 
            "Nelder-Mead", 
            "Powell"],
        "options": [
            {
                "maxiter": 100
            }
        ],
        "fun_shots": [100]
    }
}
        \end{minted}
        \captionsetup{width=0.95\linewidth}
        \caption{
            Excerpt from the descriptor file for the experiment that was performed in the context of the case study.
        }
        \label{fig:caseStudyExcerpt}   
    \end{minipage}\vspace*{-3em}
\end{figure}

\subsubsection{InitHandler}
The InitHandler plugin transpiles quantum circuits to the target back-end and initialises their parameters. Transpilation decomposes circuits into native gates, maps logical to physical qubits, and inserts SWAP operations as required by the coupling map. Parameter initialisation follows the chosen strategy.

\subsubsection{AllroundOptimizer}
The AllroundOptimizer is the core classical component of QEF’s execution pipeline, managing parameter optimisation within hybrid quantum-classical loops. It supports a wide range of gradient-free (\eg, Nelder-Mead, COBYLA, or Powell) and gradient-based (\eg, CG, BFGS, Newton-CG, L-BFGS-B, SLSQP, SGD, or ADAM) methods, while also allowing custom extensions for novel optimisation strategies.
During each iteration, it records metrics such as cost-function values, run-time, and gradients, enabling precise performance analysis. Computations use the QPU component.

\subsubsection{Quantum Processing Unit}
The QPU evaluates quantum circuits by computing both the cost function and its gradient; in Qaptiva 800, this is performed by a classical quantum emulator. QEF currently supports two such emulators—LinAlg, a tensor-based simulator, and NoisyQProc, which models device-specific noise, an essential consideration given the persistence of NISQ-era hardware. Both emulators offer reproducible results when seeded, and users may integrate custom QPUs for alternative simulators or physical devices.

\subsubsection{Post-Processing}
Depending on problem scale and optimisation method, computations may run from hours to weeks; accordingly, QEF dispatches jobs asynchronously and automatically retrieves and organises results upon completion. This non-blocking workflow, illustrated in \autoref{fig:QLMOverview}, ensures that all outputs and hyper-parameters remain traceable to their originating experiments, allowing researchers to focus on analysis rather than job management.

\subsubsection{Initialisation Strategies} 
\label{sec:InitializationStrategies}
Particularly for QAOA, QEF provides initialisation strategies proposed by Zhou~\etal~\cite{zhou:2020:physicalreviewx}, allowing researchers to tailor starting parameters to each job (\textbf{RANDOM} samples uniformly from $[0, 2\pi)$; \textbf{INTERP} initialises depth $p + 1$ by linear interpolation of depth $p$ parameters;
\textbf{FOURIER} transforms depth $p$ to the frequency domain, and seeds depth $p+1$); \textbf{Warm start} uses optimised parameters from a completed run to initialise the next, which is known to improve convergence speed~\cite{thelen:2024:qce}.

\subsection{Case Study}
To illustrate the practical benefits of QEF, we discuss an experiment on a typical combinatorial optimisation task that is solved with a quantum approach.
This shows (1) How experiments can be precisely reproduced from a descriptor file;
(2) how hypothesis-driven comparisons are supported, where a single variable can be isolated while others remain fixed; (3) the ability to generate large parameter sweeps; and (4) how results are logged in a structured and analysis-ready format
and can be post-processed with ease by using established statistical software, without making any changes to the framework itself.

We choose the seminal Max-Cut problem with QAOA, as it has been extensively
discussed in the quantum computing literature. We use an unweighted graph 
with ten nodes and a connectivity density of 0.7.

Researchers may, for instance, want to explore how different gradient-free optimisers and QAOA layer depths affect solution convergence using an experiment description like \autoref{fig:caseStudyExcerpt}.
The setup combines a standard QAOA ansatz with layers $p \in {3, 4, 5}$ and three derivative-free methods (\emph{COBYLA}, \emph{Nelder-Mead}, \emph{Powell}), yielding nine jobs that QEF executes on hardware or a simulator.
Each job varies only one parameter, pinpointing any observed differences to it.

Results per variant are stored as tidy data.
\autoref{fig:caseStudyFigures} shows the R code for \autoref{fig:caseStudyPlot}, which compares convergence across layer depths and optimisers, and \autoref{fig:caseStudyPlot2} visualises optimisation time per step.
These examples illustrate that the same data readily supports diverse metrics (\eg, approximation ratio, runtime, iteration traces) with minimal code, enabling efficient exploratory analysis.

\begin{figure}[htb]
    \centering
    \begin{subfigure}[t]{.49\textwidth}
        \input{img-tikz/convergence}
        \captionsetup{width=0.95\linewidth}
        \caption{\label{fig:caseStudyPlot}
        Convergence behaviour of QAOA in the described case study.
        }
    \end{subfigure}
    \hfill
    \begin{subfigure}[t]{.49\textwidth}
        \input{img-tikz/secondsIteration}
        \captionsetup{width=0.95\linewidth}
        \caption{\label{fig:caseStudyPlot2}
        Time per optimisation step by number of layers and optimiser.
        }       
    \end{subfigure}
    \begin{subfigure}[b]{0.9\textwidth}
        \centering
        \begin{minted}[bgcolor=LightGray]{text}
ggplot(df,aes(x=iters,y=cost_values,colour=method,linetype=factor(numLayers))) +
    geom_line() # lhs

ggplot(df,aes(x=factor(method),y=iteration_time, colour=factor(method))) +
    geom_boxplot() + scale_y_log10()+ 
    facet_wrap(facets=vars(num_layers), nrow = 1, labeller = label_value) # rhs
        \end{minted}
    \end{subfigure}
    \caption{Exploratory analysis of QAOA using QEF}
    \label{fig:caseStudyFigures}   
\end{figure}

\section{Conclusion}
The Quantum Experiment Framework (QEF) addresses a key need in hypothesis-driven quantum research by enabling reproducible, scalable experiments across both NISQ and future fault-tolerant regimes. It offers a lightweight, flexible alternative to vendor-specific tools, managing experiment orchestration, parallel execution, and automated data collection.
Our review of current practices showed that most empirical quantum studies focus on a limited set of parameters and metrics, making QEF’s targeted design~--~supporting common use cases while allowing custom extensions. By emphasising tidy data, reproducibility, and ease of use, QEF replaces ad-hoc scripting with a coherent framework that accelerates experimentation, supports rigorous analysis, and fosters comparable, community-driven evidence in quantum computing research.

\subsubsection*{Acknowledgements.}
This work was supported by the German Federal Ministry of Education and Research (BMBF), funding program ‘quantum technologies—from basic research to market’, grant number 13N16092. WM acknowledges support by the High-Tech Agenda of the Free State of Bavaria.

\renewbibmacro{in:}{}
\renewbibmacro{doi:}{}
\newbibmacro{string+doi}[1]{%
  \iffieldundef{doi}{%
  \iffieldundef{url}%
     {#1}
     {\href{\thefield{url}}{\textcolor{teal}{#1}}} 
  }%
  {\href{https://dx.doi.org/\thefield{doi}}{\textcolor{teal}{#1}}}
}
\DeclareFieldFormat*{title}{\usebibmacro{string+doi}{\mkbibemph{#1}}}

\printbibliography
\end{document}